\documentclass[12pt]{article}
\usepackage{bbold}
\usepackage{graphicx}
\def \bea{\begin{eqnarray}}
\def \beq{\begin{equation}}

\def \eea{\end{eqnarray}}
\def \eeq{\end{equation}}

\begin{document}
\rightline{EFI 11-XX}
\rightline{arXiv:1101.nnnn}
\bigskip
\centerline {\bf ANGULAR DISTRIBUTIONS IN $D_s^*$ DECAYS
\footnote{To be submitted to Phys.\ Rev.\ D}}
\bigskip
 
\centerline{Jonathan L. Rosner~\footnote{rosner@hep.uchicago.edu.}}
\centerline{Enrico Fermi Institute and Department of Physics}
\centerline{University of Chicago, 5640 S. Ellis Avenue, Chicago, IL 60637}
\medskip
 
\begin{quote}
The reaction $e^+ e^- \to (D_s^{*+} D_s^- + {\rm c.c.})$ has been used
with great success at the CLEO-c detector, and will be a source of valuable
future data from the BES-III detector, to obtain information about the
properties and decays of the $D_s$ and $D_s^*$.  The angular distributions
of the $D_s^*$ and the final-state particles in $D_s^* \to D_s \gamma$
are needed to model the decays properly, to confirm the spin-parity assignment
$J^P(D_s^*) = 1^-$, and to model the acceptance for the rare process
$D_s^* \to D_s e^+ e^-$.  This note presents some of the necessary expressions.
\end{quote}

\centerline{PACS numbers: 14.40.Lb, 13.20.Fc, 13.66.Bc, 13.40.Hq}
\bigskip

The reaction $e^+ e^-\to \gamma^* \to(D_s^{*+} D_s^-+{\rm c.c.})$ has been used
with great success at the CLEO-c detector, and will be a source of valuable
future data from the BES-III detector, to obtain information about the
properties and decays of the $D_s$ and $D_s^*$.  The angular distributions
of the $D_s^*$ and the final-state particles in $D_s^* \to D_s \gamma$
are needed to model the decays properly, to confirm the spin-parity assignment
$J^P(D_s^*) = 1^-$, and to model the acceptance for the rare process
$D_s^* \to D_s e^+ e^-$.  The purpose of the present note is to give some
of the necessary expressions for angular distributions.

We shall assume unpolarized electron-positron collisions, with the $e^+$
beam defining the $\mathbf{+ \hat z}$ direction.  The density matrix describing
the initial virtual photons $\gamma^*$ arising from $e^+ e^-$ collisions
is then proportional to the dyadic $\mathbf{\hat x \hat x + \hat y \hat y}$,
where we shall always consider linear polarization states.

Let us assume that the $D_s^{*+}$ produced opposite a $D_s^-$ (charge-conjugate
reactions are always assumed) lies in the $xz$ plane, making an angle $\theta$
with the $z$ axis.  If it were produced along the $z$ axis, its density matrix
would be of the form $\mathbf{\hat x \hat x + \hat y \hat y}$ just like that
of the virtual photon $\gamma^*$.  This may be seen from the form of the
$\gamma^* D_s^* D_s$ coupling, involving the cross product of $\gamma^*$
and $D_s^*$ polarization vectors dotted into the momentum of the $D_s^*$.
(A single power of 3-momentum is required for parity invariance.)  This
form of coupling may be seen by a short calculation to entail a $D_s^*$
production distribution of the form
\beq
W(\cos \theta) = \frac{3}{8} (1 + \cos^2 \theta)~.
\eeq
If the $D_s^*$ is produced in the $xz$ plane at an angle $\theta$ with
respect to the $z$ axis, its density matrix is now of the form $\mathbf{\hat x'
\hat x' + \hat y \hat y}$, where $\mathbf{\hat x'} \equiv \mathbf{\hat x}
\cos \theta - \mathbf{\hat z} \sin \theta$.  Thus we may write (in the $x,y,z$
basis)
\beq
\rho(D_s^*) = \left[ \begin{array}{c c c} \cos^2 \theta & 0 & -\sin \theta
\cos \theta \cr 0 & 1 & 0 \cr -\sin \theta \cos \theta & 0 & \sin^2 \theta
\end{array} \right]~.
\eeq

We now consider a photon emitted in the decay $D_s^* \to D_s^+ \gamma$.
A photon emitted along the $z$ axis can have two polarization states
$\mathbf{\epsilon}_1 = \mathbf{\hat x},~\mathbf{\epsilon}_2 = \mathbf{\hat y}$.
We shall perform an Euler rotation on the photon, first by an angle $\beta$
about the $y$ axis and then by an angle $\alpha$ about the $z$ axis.  The
photon direction then becomes
\beq
\mathbf{\hat p}_\gamma = \mathbf{\hat x} \sin \beta \cos \alpha
                        +\mathbf{\hat y} \sin \beta \sin \alpha
                        +\mathbf{\hat z} \cos \beta~,
\eeq
while the polarization vectors orthogonal to it become
\beq
\mathbf{\epsilon}'_1 = \mathbf{\hat x} \cos \beta \cos \alpha
                     + \mathbf{\hat y} \cos \beta \sin \alpha
                     - \mathbf{\hat z} \sin \beta~,
\eeq
\beq
\mathbf{\epsilon}'_2 =-\mathbf{\hat x} \sin \alpha~
                     + \mathbf{\hat y} \cos \alpha~.
\eeq
These polarization vectors are also uniquely specified by the condition
that they be orthogonal to each other and to $\mathbf{\hat p}_\gamma$, and
that $\mathbf{\epsilon}'_2$ have no $z$ component.

We now note that the decay $D_s^* \to D_s \gamma$ involves the
scalar triple product $\mathbf{\epsilon}(D_s^*) \cdot \mathbf{\epsilon}
(\gamma) \times \mathbf{p}_\gamma$, so that a photon with momentum
$\mathbf{p}_\gamma$ and no polarization component along the $z$ axis
is produced with angular distribution
\bea
W_2(\theta,\alpha,\beta) & = & \mathbf{\epsilon}'_1 \cdot \rho(D_s^{*}) \cdot
 \mathbf{\epsilon}'_1 = \cos^2 \theta \cos^2 \alpha \cos^2 \beta +
 \cos^2 \beta \sin^2 \alpha \\ & + & \sin^2 \theta \sin^2 \beta
 + 2 \sin \theta \cos \theta \sin \beta \cos \beta \cos \alpha~.
\eea
The use of $\mathbf{\epsilon}'_1$ rather than $\mathbf{\epsilon}'_2$ comes
from the scalar triple product form of the $D_s^{*+} D_s^+ \gamma$ coupling.
A photon with polarization orthogonal to the one specified above is produced
with angular distribution
\beq
W_1(\theta,\alpha,\beta) = \mathbf{\epsilon}'_2 \cdot \rho(D_s^{*}) \cdot
 \mathbf{\epsilon}'_2 = \cos^2 \alpha + \cos^2 \theta \sin^2 \alpha~.
\eeq
The sum over photon polarizations then gives a photon angular distribution
proportional to $W = W_1 + W_2$.  Special cases of the above results are
\bea
W_1(\theta = 0~{\rm or}~\pi,\alpha,\beta) & = & 1~,~~
W_2(\theta = 0~{\rm or}~\pi,\alpha,\beta) = \cos^2 \beta~,\\
W_1(\theta = \pi/2, ~\pi,\alpha,\beta) & = & \cos^2 \alpha,~
W_2(\theta = \pi/2, ~\pi,\alpha,\beta) = \cos^2 \beta \sin^2 \alpha
 + \sin^2 \beta.
\eea

If the final photon undergoes internal conversion, $\gamma \to e^+ e^-$, with
the plane of $e^+e^-$ oriented at an angle $\psi$ with respect to the
final photon polarization, the dependence on $\psi$ will be of the form
$\cos^2 \psi$.  Hence choosing as a reference polarization direction a
unit vector perpendicular to $\mathbf{p}_\gamma$ with no component in the
$\mathbf{\hat z}$ direction, the full angular distribution may be written
\beq
W(\theta,\alpha,\beta,\psi) = W_1(\theta,\alpha,\beta) \sin^2 \psi +
W_2(\theta,\alpha,\beta) \cos^2 \psi~.
\eeq

I thank Anders Ryd for asking the question leading to this investigation.
This work was supported in part by the United States Department of Energy
under Grant No.\ DE FG02 90ER40560.
\end{document}